\documentclass[amsmath,amssymb,amsfonts,aps,pre,twocolumn,superscriptaddress,showpacs,showkeys,longbibliography]{revtex4-1}

\usepackage[english]{babel}
\usepackage[utf8]{inputenc}
\usepackage[T1]{fontenc}
\usepackage{lmodern}
\usepackage{epsfig}
\usepackage{epstopdf}
\usepackage{color}
\usepackage{hyperref} 
\usepackage[caption=false,labelformat=simple]{subfig}

\makeatletter
\renewcommand{\fnum@figure}{FIG. \thefigure}
\makeatother

\definecolor{mycolor}{rgb}{0.0, 0.0, 0}

\begin{document}
\title{Spontaneous symmetry breaking of   active phase in  coevolving nonlinear voter model}
\author{Arkadiusz J\k{e}drzejewski}
\affiliation{Department of Operations Research and Business Intelligence, Wroc\l{}aw University of Science and Technology, Wybrze\.{z}e Wyspia\'{n}skiego 27, 50-370 Wroc\l{}aw, Poland}
\author{Joanna Toruniewska}
\affiliation{Center of Excellence for Complex Systems Research, Faculty of Physics, Warsaw University of Technology, ul.~Koszykowa 75, 00-662 Warsaw, Poland}
\author{Krzysztof Suchecki}
\affiliation{Center of Excellence for Complex Systems Research, Faculty of Physics, Warsaw University of Technology, ul.~Koszykowa 75, 00-662 Warsaw, Poland}
\author{Oleg Zaikin}
\affiliation{ITMO University, 49 Kronverkskiy av., 197101, Saint Petersburg, Russia}
\author{Janusz A. Ho\l{}yst}
\affiliation{Center of Excellence for Complex Systems Research, Faculty of Physics, Warsaw University of Technology, ul.~Koszykowa 75, 00-662 Warsaw, Poland}
\affiliation{ITMO University, 49 Kronverkskiy av., 197101, Saint Petersburg, Russia}

\date{\today}

\begin{abstract}
We study an adaptive network model driven by a nonlinear voter dynamics.
Each node in the network represents a voter and can be in one of two states that correspond to different opinions shared by the voters.
A voter disagreeing with its neighbor's opinion may either adopt it or rewire its link to another randomly chosen voter with any opinion. 
The system is studied by means of the pair approximation in which a distinction between the average degrees of nodes in different states is made.
This approach allows us to identify two dynamically active phases, a symmetric and an asymmetric one.
The asymmetric active phase, in contrast to the symmetric, is characterized by different numbers of nodes in the opposite states that coexist in the network.
The pair approximation predicts the possibility of spontaneous symmetry breaking, which leads to a continuous phase transition between the symmetric and the asymmetric active phases.
In this case, the absorbing transition occurs between the asymmetric active and the absorbing phases after the spontaneous symmetry breaking.
Discontinuous phase transitions and hysteresis loops between both active phases are also possible.
Interestingly, the asymmetric active phase is not displayed by the model where the rewiring occurs only to voters sharing the same opinion, studied by other authors.
Our results are backed up by Monte Carlo simulations.\\\\
Post-print of \href{https://doi.org/10.1103/PhysRevE.102.042313}{Phys. Rev. E \textbf{102}, 042313 (2020)}.\\
Copyright (2020) by the American Physical Society.
\end{abstract}

\maketitle
\section{introduction}
A feedback loop between the network topology and dynamical processes that occur between nodes is common in real-world networks \cite{Boc:etal:06,Gro:Bla:08,Say:etal:13}.
The topology impacts the evolution of node states, which in turn influence the way the structure itself is modified.
This feedback is a signature of networks that are called adaptive or coevolutionary \cite{Gro:Bla:08}.
Adaptive networks are especially relevant for social systems, where they can model phenomena such as the emergence of consensus and polarization, opinion formation, group fragmentation, or language diversity \cite{Cen:etal:07,Nar:Koz:Bar:08,Koz:Bar:08a,Rad:Gub:18}.
These coevolutionary models rely on two basic mechanisms.
One accounts for the changes in the node states, whereas the other accounts for the link rewiring.
Both of them may be implemented in various ways.
The voter model, as a minimalist model of the opinion formation process \cite{Red:19,Jed:Szn:19}, provides the basis for the evolution of state variables in many adaptive networks that represent social interactions \cite{Hol:New:06,Gil:Zan:06,Kim:Hay:08,Nar:Koz:Bar:08,Zan:Gil:06,Boh:Gro:11,Boh:Gro:12,Zsc:etal:12,Dur:etal:12,Ji:etal:13,Shi:Muc:Dur:13,Yi:etal:13,Dia:San:Egu:14,Dia:Egu:San:15,Mal:etal:16,Tor:etal:17,Kul:etal:18,Vaz:Egu:San:08,Wie:Nun:13,Mal:Muc:13,Bas:Sly:17,Cho:Muc:20}.
Other dynamics used in that context involve the nonlinear voter model \cite{Min:San:17, Rad:Min:San:18, Min:San:19,Kur:Por:19,Rad:San:20}, the Deffuant model \cite{Koz:Bar:08a, Koz:Bar:08b}, the Axelrod model \cite{Rad:Gub:17,Rad:Gub:18,Cen:etal:07}, or the $q$-state Potts model \cite{Tor:Suc:Hol:16}.
Interactions between nodes can also be defined by a Hamiltonian that depends on topological properties of a social network \cite{Rad:etal:18}.

When it comes to the link rewiring mechanisms, most of them reflect the effect known in sociology as homophily, which is the tendency of individuals to bond with others who are similar to themselves  \cite{Mcp:Smi:Coo:01,Cen:etal:07}.
Under this paradigm, nodes may remove their links to disagreeing neighbors and form new ones to randomly chosen nodes in the same states \cite{Kul:etal:18,Boh:Gro:12,Zsc:etal:12,Kim:Hay:08,Hol:New:06,Ji:etal:13,Dia:Egu:San:15,Min:San:19,Dia:San:Egu:14,Rad:Min:San:18,Dur:etal:12,Kur:Por:19,Tor:etal:17,Boh:Gro:11,Min:San:17,Vaz:Egu:San:08,Wie:Nun:13,Bas:Sly:17,Cho:Muc:20,Rad:San:20}.
Heterophily, as the opposite effect to homophily, is modeled as a preference to connect to individuals with distinct traits \cite{Kim:Hay:08}.
Another approach is not to distinguish between states at all so that links can be rewired to any nodes of the network \cite{Koz:Bar:08b,Nar:Koz:Bar:08,Koz:Bar:08a,Mal:etal:16,Shi:Muc:Dur:13,Dur:etal:12,Kur:Por:19,Bas:Sly:17,Cho:Muc:20}.
Additional modifications such as link removal \cite{Zan:Gil:06,Gil:Zan:06,Kur:Por:19}, triadic closure \cite{Mal:etal:16,Rad:Min:San:18,Rad:Gub:17,Rad:Gub:18,Mal:Muc:13}, or different preferential attachment schemes \cite{Rad:Gub:17,Rad:Gub:18,Mal:Muc:13} are considered as well in order to capture some properties of real networks.

The competition between these two mechanisms, which are responsible for the changes in the node states and the network structure, in adaptive networks leads frequently to a fragmentation transition, where the network splits into smaller components.
One of the simplest models that displays this kind of behavior is a coevolving voter model \cite{Vaz:Egu:San:08}.
Being analytically tractable, it has played a fundamental role in understanding the process of network fragmentation \cite{Vaz:Egu:San:08,Boh:Gro:11}.
This work extends the study in this area via an analysis of one of its nonlinear extensions.

In Ref.~\cite{Dur:etal:12}, two coevolving voter models that are different only in the rewiring mechanisms were compared.
In the model with the rewire-to-same mechanism, new links can be established only between nodes in the same states, as in Ref.~\cite{Vaz:Egu:San:08}, whereas with the rewire-to-random mechanism, new links can be established between all nodes regardless of their states, as in Ref.~\cite{Nar:Koz:Bar:08}.
This small difference in the dynamics led to different transition types exhibited by the models in finite systems \cite{Dur:etal:12}.
However, later on, more research attention was directed towards the model with the rewire-to-same mechanism, in which the role of nonlinear interactions between voters has been studied on single-layer \cite{Min:San:17} and two-layer \cite{Min:San:19} networks. 
The introduction of this kind of nonlinearity into the model resulted in the appearance of new phases and fragmentation transitions \cite{Min:San:17,Min:San:19}.
In this regard, the analysis of the nonlinear version of the coevolving voter model with the rewire-to-random mechanism seems to be interesting not only for comparative but also cognitive reasons since it may potentially reveal some other phenomena related to the network fragmentation.
In this work, we carry out such an analysis.

The coevolving nonlinear voter model with the rewire-to-random mechanism is studied by means of the pair approximation in which we distinguish between the average degrees of nodes in different states.
In the analysis of adaptive systems, it is important to make such a distinction. 
This is because the feedback loop between the node states and the network structure makes these average node degrees different from each other in general \cite{Nar:Koz:Bar:08, Tor:etal:17, Wie:Nun:13}.
Since this formalism allows for more accurate model characterization, it may also expose some additional properties of the system.
In fact, it has already contributed to the discovery of a non-trivial conservation law in the coevolving voter model with the rewire-to-same mechanism \cite{Tor:etal:17}.
Nevertheless, when it comes to the nonlinear extensions of coevolving voter models, none of them has been studied within this approach so far.

In coevolving nonlinear voter models, the degree of nonlinearity is measured by the parameter $q$, which determines the functional form of interaction probabilities between nodes \cite{Min:San:17,Min:San:19,Rad:Min:San:18,Kur:Por:19}.
The same kind of nonlinearity has been considered in various nonlinear $q$-voter models on static structures \cite{Cas:Mun:Pas:09,Nyc:Szn:Cis:12,Chm:Szn:15,Jed:17,Per:etal:18}; a more extensive review can be found in Ref.~\cite{Jed:Szn:19}.
For values $0<q\leq1$, our calculations reveal qualitative differences between the coevolving nonlinear voter model with the rewire-to-random mechanism, studied herein, and its rewire-to-same counterpart, studied in Ref.~\cite{Min:San:17}. 
Therefore, we focus our analysis on this specific range of the parameter.

In the thermodynamic limit, the fragmentation transition in the coevolving nonlinear voter model with the rewire-to-same mechanism occurs between dynamically active and absorbing phases  \cite{Min:San:17}.
The active phase is characterized by the presence of active links that connect nodes in different states and drive the dynamics. 
In this phase, the network remains connected; however, its structure and the states of the nodes are constantly changing.
In the absorbing phase, there are no active links, and the network splits into two components, each of which is composed of nodes in the same state.
In the model with the rewire-to-same mechanism, the active phase is symmetric in the sense that the stationary numbers of nodes in different states are equal, so neither of the states is preferred in the network.
Interestingly, in the model with the rewire-to-random mechanism, these numbers can also be different, so we can identify the asymmetric active phase as well. 
The asymmetric active phase is characterized by a predominance of nodes in one state so that this state is preferred in the network.
For a specific range of the nonlinearity parameter, i.e., for $q^*\leq q<1$, where $q^*=\frac{1}{6} (\sqrt{13}+1) \approx0.7676$, the pair approximation predicts spontaneous symmetry breaking and a continuous phase transition between the symmetric and the asymmetric active phases.
We characterize the critical properties of this transition.
Discontinuous phase transitions between both active phases are also possible, and they appear for $q<q^*$.

In this work, we analyze a rich phase diagram displayed by the coevolving nonlinear voter model with the rewire-to-random mechanism on the pair approximation level.
The presence of the asymmetric active phase, identified by our analysis, is confirmed by Monte Carlo simulations.

\section{model description}
\label{sec:model}
\begin{figure}[b!]
	\centering
	\epsfig{file=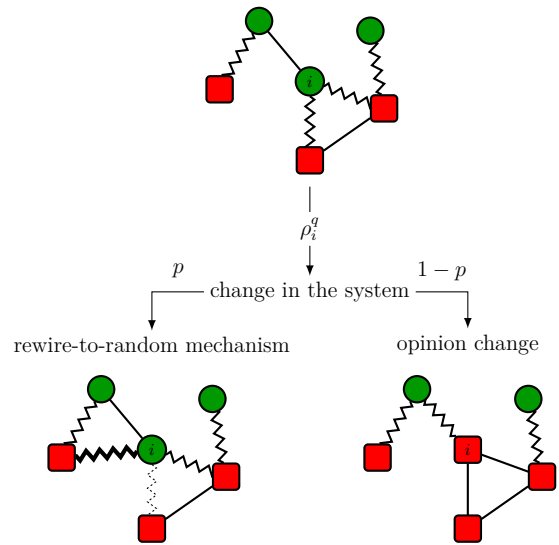}
	\caption{\label{fig:model} Schematic illustration of one update of the coevolving nonlinear voter model with the rewire-to-random mechanics.
		Node shapes symbolize opinions, while zigzag and straight lines refer to active and inactive links, respectively.
		In this example, node $i$ is chosen randomly, so the interactions with its neighbors cause a change in the system with probability $\rho_i^q$.
		In case of the change, the node $i$ breaks its one randomly chosen active link (a dotted zigzag in the figure) and establishes a new link (a thick zigzag  in the figure) to a randomly chosen node with probability $p$, or it changes its opinion to the opposite one with probability $1-p$.
	}
\end{figure}
\begin{figure*}[t!]
	\centering
	\epsfig{file=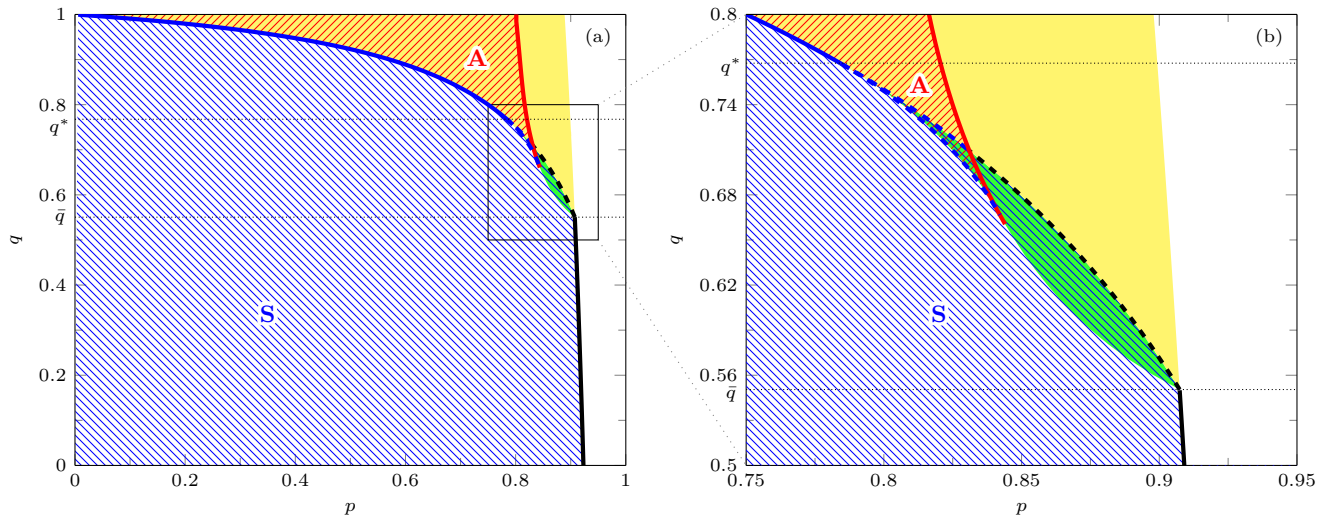}
	\caption{\label{fig:phase-diagram-pq-k6} Phase diagrams for the network with $\langle k\rangle=6$. \textcolor{mycolor}{The areas with blue and red stripes}  correspond to the stable solutions associated with the symmetric ($S$) and the asymmetric ($A$) active phases, respectively.
	On the other hand, the \textcolor{mycolor}{yellow} and \textcolor{mycolor}{green} areas correspond to the unstable solutions associated with these phases (\textcolor{mycolor}{yellow} for the symmetric and \textcolor{mycolor}{green} for the asymmetric phase).
	\textcolor{mycolor}{
	In the areas without any stripes, only the absorbing phase is stable.
	The solid and dashed thick lines indicate continuous and discontinuous phase transitions, respectively, whereas the line colors indicate between which phases the transition occurs: blue, between the symmetric and the asymmetric active phase; red, between the asymmetric active and the absorbing phase; and black, between the symmetric active and the absorbing phase.}
    For more details, see the text. See as well the representative stability diagrams projected onto the $(\rho, p)$ plane that are presented in Fig.~\ref{fig:sample-phase-diags} for several values of $q$, 
    \textcolor{mycolor}{where all these transitions can be observed directly}.
	}
\end{figure*}

We consider an undirected network comprised of $N$ nodes, which represents a social structure.
Nodes stand for voters, and each of them has an opinion that is expressed as a variable $j\in\{1,-1\}$, or equivalently $j\in\{\uparrow,\downarrow\}$ for simplicity of notation. Links indicate the mutual influence of voters on each other's opinion. 
Randomly, one node is selected after another.
Let $\rho_i$ denote the concentration of disagreeing neighbors with the selected node $i$.
Formally, $\rho_i=a_i/k_i$, where $a_i$ is the number of active links attached to the node $i$, and $k_i$ is its degree.
In other words, $\rho_i$ is the local concentration of active links.
With probability $\rho_i^q$, the interactions between the node $i$  and its neighbors cause a change in the system, whereas with complementary probability $1-\rho_i^q$, nothing happens.
The nonlinearity parameter, $q$, is in the range $q>0$.
In case of the change, two events are possible.
With probability $p$, one randomly picked active link of the node $i$ is rewired to another node picked at random from all the nodes in the network. Otherwise, with probability $1-p$, the node $i$ changes its opinion to the opposite.
Figure~\ref{fig:model} schematically illustrates one update of the above dynamics.
One time step is understood as $N$ such updates.

The only difference between this model and the model analyzed in Ref.~\cite{Min:San:17} is that the model from the reference adopts the rewire-to-same mechanism instead of the rewire-to-random mechanism adopted herein. We show that this difference is important since the rewire-to-random mechanism makes possible the emergence of spontaneous symmetry breaking and the emergence of the asymmetric active phase, which is absent in the model with the second mechanism.
The linear versions of both the models, which correspond to $q=1$, are compared in Refs.~\cite{Dur:etal:12,Bas:Sly:17}.
Both the mechanisms are also analyzed in the adaptive voter model with noise in Ref.~\cite{Cho:Muc:20}.
However, these linear models implement a link-based updating scheme in contrast to the nonlinear models studied herein and in Ref.~\cite{Min:San:17}, which implement a node-based updating scheme.

\section{pair approximation result discussion}
\begin{figure*}[t!]
	\centering
	\epsfig{file=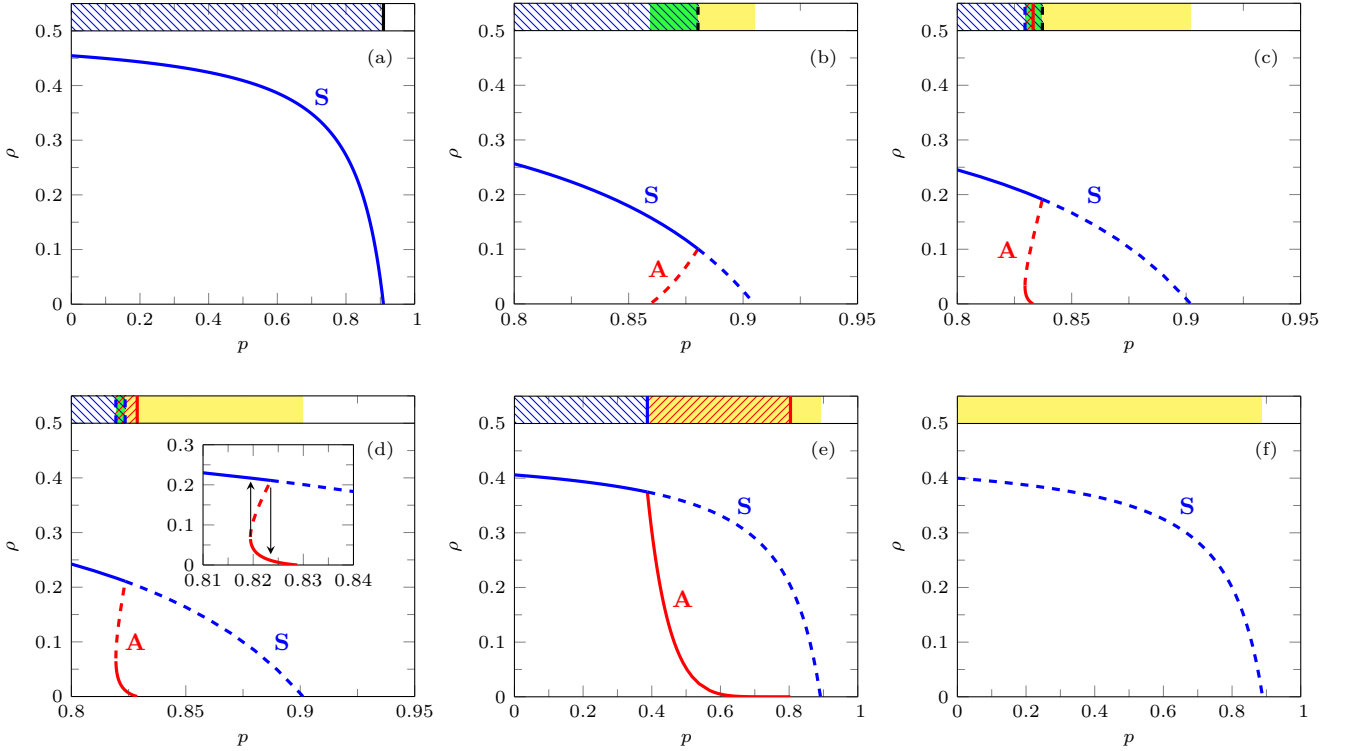}
	\subfloat{\label{fig:pda-a}}
	\subfloat{\label{fig:pda-b}}
	\subfloat{\label{fig:pda-c}}
	\subfloat{\label{fig:pda-d}}
	\subfloat{\label{fig:pda-e}}
	\subfloat{\label{fig:pda-f}}
	\caption{\label{fig:sample-phase-diags} Representative stability diagrams for the network with $\langle k\rangle=6$ and different values of $q$: (a) $0<q<\bar{q}$, (b)-(d) $\bar{q}<q<q^*$, (e) $q^*<q<1$, and (f) $q=1$, where $q^*\approx0.7676$ and $\bar{q}\approx0.5505$ for the given average node degree of the network.
	Part~\ref{fig:pda-d} contains an inset with a magnified region where a discontinuous phase transition between the symmetric and the asymmetric active phase occurs. The hysteresis loop is indicated by arrows.
	The solid and dashed lines correspond to the stable and unstable solutions, respectively. The \textcolor{mycolor}{blue} lines refer to the symmetric ($S$) active phase (for which $c=0.5$ and $m=0$), whereas the \textcolor{mycolor}{red} lines refer to the asymmetric ($A$) active phase (for which $c\neq0.5$ and $m\neq0$). 
	The exact values of $q$ in the plots are as follows: (a) $q=0.5$, (b) $q=0.62$, (c) $q=0.7$, (d) $q=0.72$, and (e) $q=0.95$.
	On top of each diagram, a slice of Fig.~\ref{fig:phase-diagram-pq-k6} for the corresponding parameter $q$ is shown.}
\end{figure*}

The system is described by three state variables: the concentration of nodes with the opinion $j=1$, the concentration of active links, and the link magnetization, denoted by $c$, $\rho$, and $m$, respectively.
\textcolor{mycolor}{Additionally, let $\langle k\rangle$ denote the average node degree calculated for the whole network, and $\langle k_j\rangle$ is the average node degree calculated only for the nodes in the state $j$.
Explicit definitions and differential equations that set the time evolution of the state variables can be found in the Appendix~\ref{app:pairapp} together with details of the used approximation.
Herein, we focus just on the steady solutions of said equations as they correspond to different phases, the discussion of which is the core of this paper.
}

\begin{figure*}[t!]
	\centering
	\epsfig{file=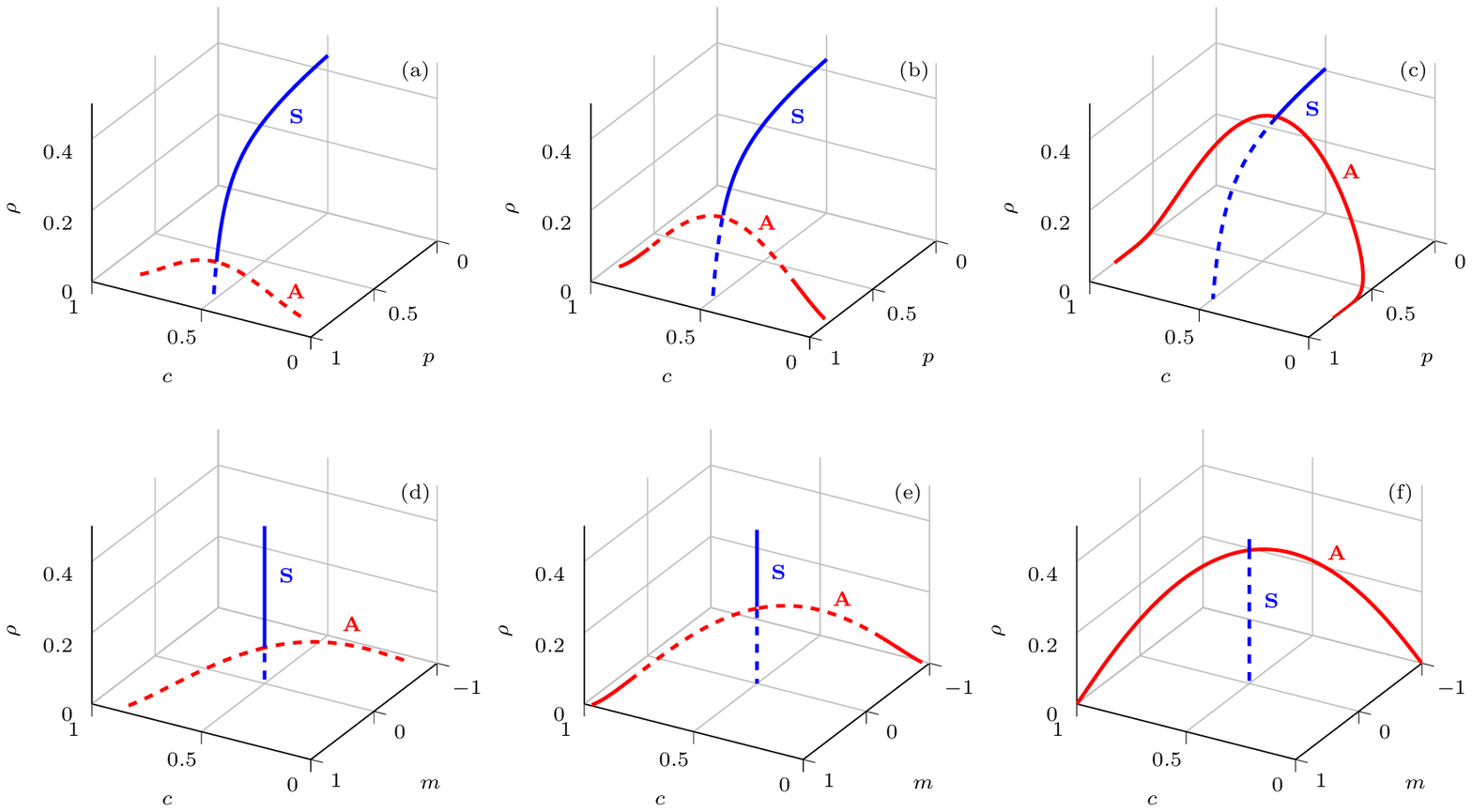}
	\subfloat{\label{fig:fpd-a}}
	\subfloat{\label{fig:fpd-b}}
	\subfloat{\label{fig:fpd-c}}
	\subfloat{\label{fig:fpd-d}}
	\subfloat{\label{fig:fpd-e}}
	\subfloat{\label{fig:fpd-f}}
	\caption{\label{fig:phase3d} Stability diagrams for the network with $\langle k\rangle=6$ depicted in different spaces: $(c,p,\rho)$ in the top and $(c,m,\rho)$ in the bottom row. Each column corresponds to one value of the nonlinearity parameter: $q=0.62$, $q=0.72$, and $q=0.95$ from left to right. The same parameters are used in Figs.~\ref{fig:pda-b}, \ref{fig:pda-d}, and \ref{fig:pda-e}, which are the projections of the diagrams in the top row to a two-dimensional space $(p,\rho)$. The solid and dashed lines correspond to the stable and unstable solutions, respectively. The \textcolor{mycolor}{blue} lines refer to the symmetric ($S$) active phase (for which $c=0.5$ and $m=0$), whereas the \textcolor{mycolor}{red} lines  refer to the asymmetric ($A$) active phase (for which $c\neq0.5$ and $m\neq0$). }
\end{figure*}

\textcolor{mycolor}{Throughout the work, we distinguish between three phases---the absorbing phase, the symmetric active phase, and the asymmetric active phase. The absorbing phase is characterized by $\rho=0$.}
In this case, the final values of $c$ and $m$ are determined by the initial conditions.
The absorbing phase is dynamically inactive, i.e., there are no further changes in the network structure nor in the voters' opinions.
On the other hand, if $\rho>0$, we have a phase that is dynamically active, and the network together with the opinions are constantly changing.
In this phase, the steady values of $c$, $\rho$, and $m$, fulfill the following equations:
\begin{equation}
m=\frac{\sqrt[q]{c}-\sqrt[q]{1-c}}{\sqrt[q]{c}+\sqrt[q]{1-c}},
\label{eq:activeM}
\end{equation}
\begin{equation}
p=\frac{[m-2c+1]\langle k\rangle}{m\langle k\rangle+(2c-1)\left[2c(1-c)-\langle k\rangle\right]},
\label{eq:activeP}
\end{equation}
and
\begin{equation}
2\rho\left[\frac{\langle k_\uparrow\rangle-q}{1+m}+\frac{\langle k_\downarrow\rangle-q}{1-m}\right]=\langle k_\uparrow\rangle+\langle k_\downarrow\rangle-4q-\frac{p}{1-p}.
\label{eq:activeR}
\end{equation}
In our model, we can distinguish between two active phases based on the steady values of $c$ and $m$.
The symmetric active phase corresponds to $c=1/2$ and $m=0$, whereas the asymmetric active phase corresponds to $c\neq1/2$ and $m\neq0$.
In contrast, the coevolving nonlinear voter model with the rewire-to-same mechanism does not exhibit the asymmetric active phase \cite{Min:San:17}.
Note that in the active phases,  $c=1/2$ implies $m=0$, and $c>1/2$ ($c<1/2$) when $m>0$ ($m<0$) based on Eq.~(\ref{eq:activeM}) when $q>0$. This means that the group of nodes that hold the majority opinion (which we understand to be $j=1$ when $c>1/2$, and $j=-1$ when $c<1/2$) has more links connecting voters with the same opinions, $\uparrow\uparrow$ or  $\downarrow\downarrow$, than the group with the minority opinion.
Moreover, having combined Eqs.~(\ref{eq:kup}), (\ref{eq:kdown}), and (\ref{eq:activeM}), we get
\begin{equation}
    \langle k_\uparrow\rangle=\langle k_\downarrow\rangle\left(\frac{1+m}{1-m}\right)^{1-q}.
\end{equation}
Thus, $\langle k_\uparrow\rangle=\langle k_\downarrow\rangle=\langle k\rangle$ only in the symmetric active phase since then $m=0$.
Note that $\langle k_\uparrow\rangle>\langle k_\downarrow\rangle$ ($\langle k_\uparrow\rangle<\langle k_\downarrow\rangle$) when $m>0$ ($m<0$) for the values of the nonlinearity parameter that we consider, i.e., $0<q<1$.
This means that nodes that hold the majority opinion have on average higher degrees when $0<q<1$.

Figure~\ref{fig:phase-diagram-pq-k6} illustrates a phase diagram for our model placed on the network with the average node degree ${\langle k\rangle=6}$.
In the diagram, the regions marked by the blue and the red \textcolor{mycolor}{stripes} correspond to the stable solutions associated with the symmetric and the asymmetric active phases, respectively. In contrast, the unstable solutions associated with these phases are depicted by the \textcolor{mycolor}{yellow} and \textcolor{mycolor}{green regions} (\textcolor{mycolor}{yellow} for the symmetric and \textcolor{mycolor}{green} for the asymmetric phase).
\textcolor{mycolor}{The solid and the dashed lines, on the other hand, indicate continuous and discontinuous phase transitions, respectively.}
The stability was checked numerically by the linearization technique.

For $0<q<\bar{q}$, where
\begin{equation}
	\bar{q}=\frac{1}{2}\left[\langle k\rangle-\sqrt{\langle k\rangle^2-2\langle k\rangle}\right],
\end{equation}
all the steady solutions are stable and are situated on a curve in the space $(c,\rho,m)$ for which $c=1/2$,
\begin{equation}
\rho=\frac{2(1-p)(\langle k\rangle-2q)-p}{4(1-p)(\langle k\rangle -q)},
\label{eq:rhom0}
\end{equation}
and $m=0$.
In this case, the system displays a continuous phase transition between the symmetric active and the absorbing phase at
\begin{equation}
	p^*=\frac{\langle k\rangle-2q}{\langle k\rangle-2q+1/2},
	\label{eq:pstar}
\end{equation} where the concentration of active links, $\rho$, is an order parameter, like in other similar models \cite{Vaz:Egu:San:08,Min:San:17}. 
In Fig.~\ref{fig:phase-diagram-pq-k6}, the solid black line corresponds to $p^*$.
For $p<p^*$, $\rho>0$, and the system is in the symmetric active phase, where nodes in different states coexist and form groups of equal sizes (since $c=1/2$).
Along with the increasing control parameter $p$, the concentration of active links, $\rho$, continuously decreases and becomes zero at $p^*$.
In the vicinity of the critical point, $p^*$, we can approximate Eq.~(\ref{eq:rhom0}) by
\begin{equation}
\rho=\frac{p^*-p}{4(\langle k\rangle-q)(p^*-1)^2}.
\end{equation}
Thus, the critical exponent associated with the order parameter in this case is $\beta=1$ since $\rho\sim(p^*-p)^\beta$, just like for the coevolving voter model in Ref.~\cite{Vaz:Egu:San:08}.
For $p>p^*$, the system dynamics ends in the absorbing phase, for which $\rho=0$; see Fig.~\ref{fig:pda-a}.
This type of a transition is also displayed by the coevolving nonlinear voter model with the rewire-to-same mechanism for $0<q\leq1$ \cite{Min:San:17} or by its linear predecessor \cite{Vaz:Egu:San:08}.

On the other hand, the system behavior for $\bar{q}<q\leq1$ is more complex and different from that exhibited by the coevolving nonlinear voter model with the rewire-to-same mechanism \cite{Min:San:17}.
First of all, the steady solutions associated with the symmetric active phase [for which $\rho$ is given by Eq.~(\ref{eq:rhom0}), $c=1/2$, and $m=0$] are stable for $p<p_s$, where
\begin{equation}
p_s=\frac{2\langle k\rangle(1-q)}{2\langle k\rangle(1-q)+q}.
\label{eq:ps}
\end{equation}
Otherwise, these solutions are unstable; see Figs.~\ref{fig:pda-b}-\ref{fig:pda-f}.
\textcolor{mycolor}{In Fig.~\ref{fig:phase-diagram-pq-k6}, the boundary between  the regions marked by blue stripes and yellow color corresponds to $p_s$.}

Secondly, there are also steady solutions given by Eqs.~(\ref{eq:activeM})-(\ref{eq:activeR}) for which $\rho>0$, $c\neq1/2$, and $m\neq0$.
These solutions correspond to the asymmetric active phase, where nodes in different states coexist and form groups of different sizes (since $c\neq1/2$).
For $\bar{q}<q<q^*$, where $q^*=\frac{1}{6}(\sqrt{13}+1)\approx0.7676$, these solutions may be either stable or unstable \textcolor{mycolor}{(the regions marked by red stripes and green color, respectively, in Fig.~\ref{fig:phase-diagram-pq-k6}).} 
In contrast to $\bar{q}$, $q^*$ does not depend on the average node degree of the network.
In this region, discontinuous phase transitions are possible. \textcolor{mycolor}{ They are marked by dashed lines in Fig.~\ref{fig:phase-diagram-pq-k6}}.
These discontinuous transitions may occur between the symmetric active and the absorbing phase, see Figs.~\ref{fig:pda-b} and \ref{fig:pda-c}, or directly between both active phases, see Figs.~\ref{fig:pda-c} and \ref{fig:pda-d}.
\textcolor{mycolor}{Note the inset in Fig.~\ref{fig:pda-d} with the hysteresis loop indicated by arrows.
The region where the system is bistable corresponds to a small triangular part of the diagram in Fig.~\ref{fig:phase-diagram-pq-k6} constrained by two dashed blue lines and a solid red line.
The stability diagrams depicted in $(p, \rho)$ space in Figs.~\ref{fig:pda-b} and \ref{fig:pda-d} are also presented in Figs.~\ref{fig:fpd-a}, \ref{fig:fpd-b}, \ref{fig:fpd-d}, and \ref{fig:fpd-e}, however, in $(c,p,\rho)$ and $(c,m,\rho)$ spaces. 
We do not present the diagram from Fig.~\ref{fig:pda-c} in these spaces since it would look similar to the diagrams in Figs.~\ref{fig:fpd-b} and \ref{fig:fpd-e} for the parameter ranges covered in these figures.}


For $q^*\leq q<1$, the solutions that correspond to the asymmetric active phase are always stable; \textcolor{mycolor}{see Fig.~\ref{fig:phase-diagram-pq-k6}}.
In this case, the absorbing transition occurs between the asymmetric active and the absorbing phase after spontaneous symmetry breaking in the active phase at $p_s$; see Fig.~\ref{fig:pda-e} or Figs.~\ref{fig:fpd-c} and \ref{fig:fpd-f}.
\textcolor{mycolor}{In Fig.~\ref{fig:phase-diagram-pq-k6}, the solid red line corresponds to the absorbing transition, whereas the solid blue line corresponds to the spontaneous symmetry breaking.}
In the vicinity of $p_s$, we can analyze the critical behavior of all our state variables.
Let us start with the link magnetization.
For $p<p_s$, the only stable solution is associated with $m=0$. However, when $p>p_s$, we can write down the following expansion, which is fulfilled by two stable values of $m$ for a given value of $p$:
\begin{equation}
p-p_s=q(q-q^*)p_s^2\frac{6q+\sqrt{13}-1}{12(1-q)\langle k\rangle}m^2+\mathcal{O}(m^4).
\end{equation}
Therefore, the critical exponent associated with the link magnetization is $\beta=1/2$ for $q^*<q<1$, and $\beta=1/4$ for $q=q^*$ since then the first coefficient that does not disappear in the expansion stands next to $m^4$.
Similar critical behavior is displayed by the node magnetization (defined as $2c-1$) since near $p_s$, $2c-1=qm$, i.e., the node magnetization is proportional to the link magnetization.
Finally, we can associate two critical exponents with the concentration of active links depending on the active phase in which we approach $p_s$; see Figs.~\ref{fig:pda-e} or \ref{fig:fpd-c}.
Let us call them $\beta_S$ and $\beta_A$ for the symmetric and the asymmetric active phase, respectively.
Thus, we have $\rho-\rho_s\sim(p_s-p)^{\beta_S}$ for $p<p_s$, and $\rho_s-\rho\sim(p-p_s)^{\beta_A}$ for $p>p_s$, where $\rho_s$ is the value of the concentration of active links at the point of spontaneous symmetry breaking $p_s$, i.e.,
\begin{equation}
\rho_s=\frac{2q(\langle k\rangle-q)-\langle k\rangle}{2q(\langle k\rangle-q)}.
\end{equation}
Having conducted the series expansion of $\rho$ at $p_s$, we get that $\beta_S=1$ for $q^*\leq q<1$, whereas $\beta_A=1$ for $q^*<q<1$, and $\beta_A=1/2$ for $q=q^*$.
Therefore, the system has different critical exponents on both sides of the transition for $q=q^*$.

For $q=1$, the asymmetric active phase disappears, whereas the symmetric active phase is unstable; see Fig.~\ref{fig:pda-f}.

\section{Monte Carlo simulations}
\begin{figure*}[t]
	\centering
	\epsfig{file=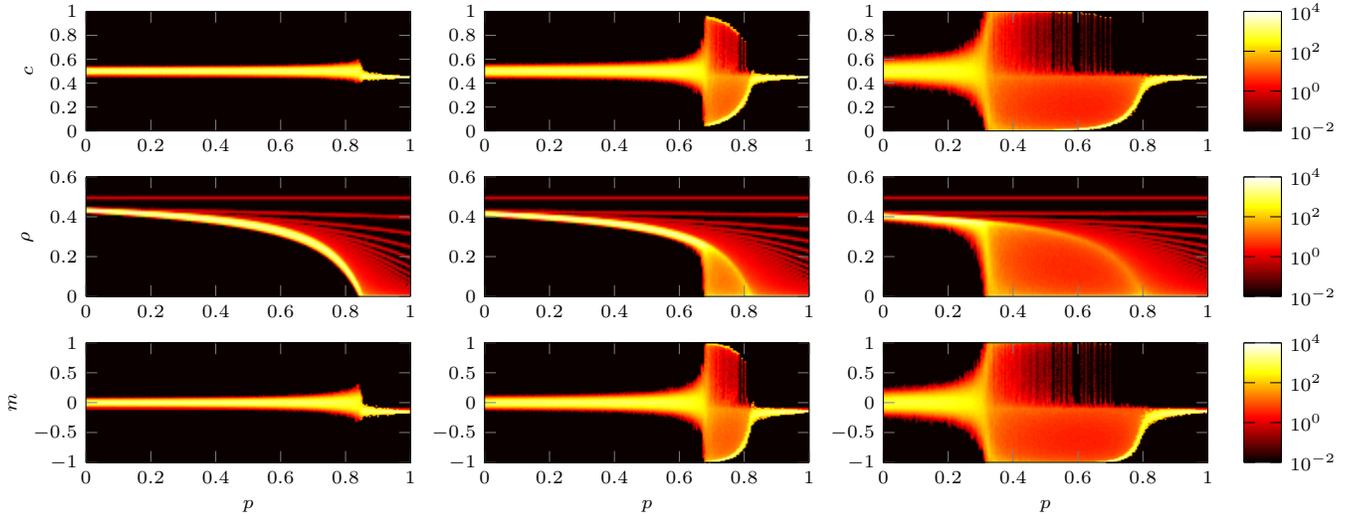}
	\caption{\label{fig:heatmapMC} Heatmaps that represent the average time spent by the system in a given state during its time evolution with the horizon 5000 time steps. 
		The data comes from Monte Carlo simulations of the network with $N=10^4$ nodes and $\langle k\rangle=6$.
		The simulations start from a random distribution of opinions so that on average $c_0=0.45$. 
		The results are averaged over 100 realizations.
		Each column in the figure corresponds to one value of the nonlinearity parameter: $q=0.6$, $q=0.8$, and $q=0.95$ from left to right \textcolor{mycolor}{(we tried to choose such values of $q$ that qualitatively reproduce the theoretical heatmaps presented in Fig.~\ref{fig:heatmapAnal})}.
		For $q=0.95$, there is a range of $p$ for which the system stays neither in the symmetric active phase nor in the absorbing phase. This region corresponds to the asymmetric active phase.
	}
\end{figure*}
\begin{figure*}[th!]
	\centering
	\epsfig{file=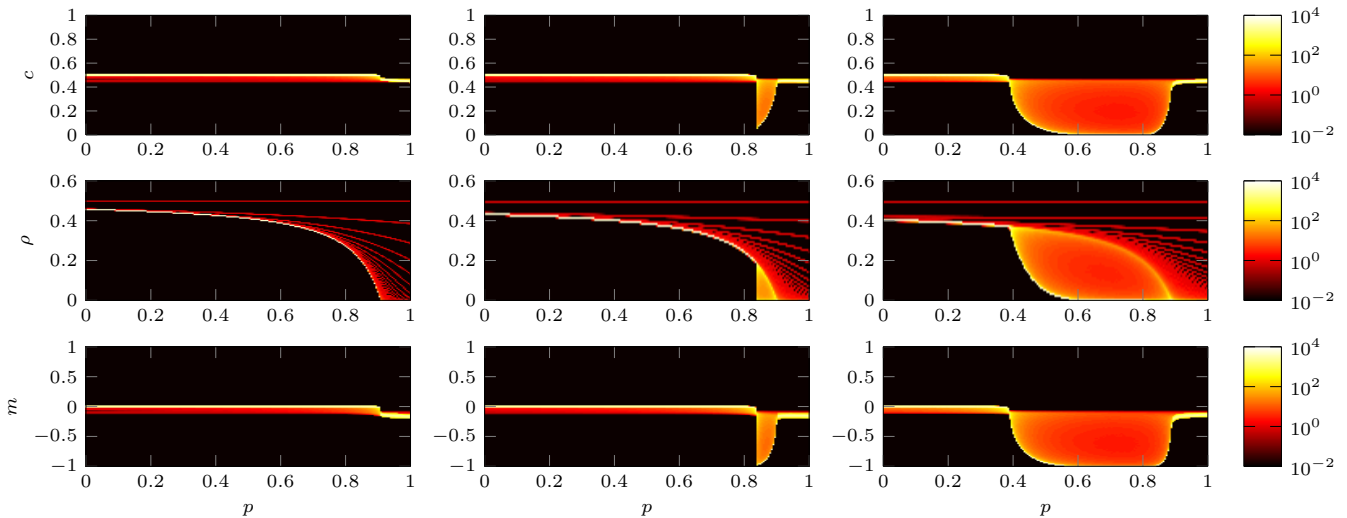}
	\caption{\label{fig:heatmapAnal} Heatmaps that represent the average time spent by the system in a given state during its time evolution with the horizon 5000 time steps. 
		Used trajectories come from the pair approximation, i.e., the numerical solutions of Eqs.~(\ref{eq:dc})-(\ref{eq:dm}). As in Fig.~\ref{fig:heatmapMC}, $\langle k\rangle=6$, and the random initial conditions are used so that on average $c_0=0.45$.
		The results are averaged over 100 trajectories.
		Each column in the figure corresponds to one value of the nonlinearity parameter: \textcolor{mycolor}{$q=0.5$ ($q<\bar{q}$), $q=0.7$ ($\bar{q}<q<q^*$), and $q=0.95$ ($q>q^*$)} from left to right.
		\textcolor{mycolor}{For $q=0.5$,  the system displays a continuous phase transition between the symmetric active and the absorbing phase. For $q=0.7$, a discontinuous phase transition between the symmetric active and the absorbing phase occurs. For $q=0.95$, we have two continuous phase transitions---the first one between the symmetric and the asymmetric active phase and the second one between the asymmetric active and the absorbing phase. In the simulations, the asymmetric active phase is much narrower (see the last column of Fig.~\ref{fig:heatmapMC}).}}
\end{figure*}
Although our analytical calculations rely on some approximations, the existence of the asymmetric active phase in the coevolving nonlinear voter model with the rewire-to-random mechanism is confirmed by Monte Carlo simulations.
The model starts its time evolution on the Erd\"{o}s-R\'{e}nyi network \cite{Alb:Bar:02} with $N=10^4$ nodes and the average node degree $\langle k\rangle=6$.
At the beginning, the opinions are randomly distributed among nodes in a way that on average gives $c_0=0.45$ ($c_0$ is an expected value, not a sample mean).
Due to fluctuations in finite systems, such a model always eventually reaches the absorbing phase.
Thus, in order to detect the active phases, we use heatmaps that represent the average time spent by the system in a given state during its time evolution.
The time horizon of our simulations amounts to $5000$ time steps, and the results are averaged over $100$ realizations.
Figure~\ref{fig:heatmapMC} illustrates such heatmaps for three different values of the nonlinearity parameter $q$, one for each column.
On the other hand, Fig.~\ref{fig:heatmapAnal} presents theoretical heatmaps depicted based on the numerical solutions of \textcolor{mycolor}{the equations that set the time evolution of the system, derived based on the pair approximation, i.e., Eqs.~(\ref{eq:dc})-(\ref{eq:dm})}.
Note that these equations are for the average values of the state variables. Thus, they do not account for the fluctuations that occur during the system dynamics in the simulations. 
However,  we took into account the fluctuations connected with the initial distribution of opinions in the simulations by solving Eqs.~(\ref{eq:dc})-(\ref{eq:dm}) from different initial conditions and then averaging the results.
Thus, each numerical trajectory, used for creating Fig.~\ref{fig:heatmapAnal}, starts from $c$  that comes from the same distribution as the sample average of the initial values of $c$ in the simulations (i.e., the distribution of $X/N$, where the number  $X$ of  nodes  with the initial opinion $j = 1$ follows the binomial distribution with parameters $N$ and $c_0$).
In the heatmaps from  Monte Carlo simulations, we see that the system sometimes passes through the states for which $c>0.5$ although it does not happen in the analytical heatmaps. 
This is because of the fluctuations that occur during the system dynamics, which are present only in Monte Carlo simulations. 
For the chosen parameters, the fluctuations connected with the initial conditions for the solutions of the pair approximation are too small to make the system pass through $c>0.5$ when the evolution starts from $c_0=0.45$ (then the theoretical standard deviation of the initial $c$ is around $0.005$).

\begin{figure}[!t]
	\centering
	\epsfig{file=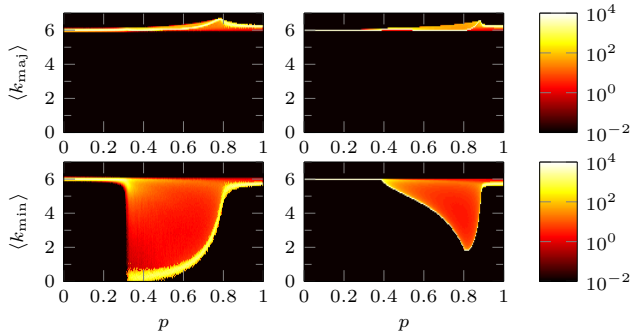}
	\caption{\label{fig:heatmapkave} Heatmaps that represent the average time spent by the system with a given average node degree during its time evolution with the horizon 5000 time steps. 
	The average node degrees, $\langle k_{\text{maj}}\rangle$ and $\langle k_{\text{min}}\rangle$,  are calculated among nodes that hold the majority and the minority opinions, respectively.
	The first column refers to Monte Carlo simulations, whereas the second one to the outcomes of the pair approximation. As in the previous heatmaps, $\langle k\rangle=6$, and the random initial conditions are used so that on average $c_0=0.45$. The results are averaged over 100 realizations, and $q=0.95$.}
\end{figure}

\textcolor{mycolor}{Having compared corresponding columns in Figs.~\ref{fig:heatmapMC} and \ref{fig:heatmapAnal}, we see qualitative similarities between the results from the simulations and the pair approximation. 
We tried to choose such values of $q$ in the simulations that qualitatively reproduce the theoretical heatmaps depicted for three characteristic ranges of the nonlinearity parameter.
Note that the exact values of $\bar{q}$ and $q^*$ in the simulations may differ from those derived based on the pair approximation. 
In Fig.~\ref{fig:heatmapAnal}, the first column captures a continuous phase transition between the symmetric active and the absorbing phase, which occurs for $q<\bar{q}$.
The second column corresponds to the case $\bar{q}<q<q^*$, where discontinuous phase transitions are possible.
The symmetric active phase loses its stability at some point, and a discontinuous phase transition occurs to the absorbing phase.
In fact, due to absorbing nature of the model, it is difficult to state unambiguously whether in the simulation, we have a discontinuous phase transition to the absorbing phase or to the asymmetric active phase that is narrow and close to the absorbing one.
The last column illustrates a continuous phase transition between the symmetric and the asymmetric active phase, which occurs for $q>q^*$.
After the symmetry of the active phase is spontaneously broken, a second continuous phase transition  takes place but this time from the asymmetric active to the absorbing phase.}
As seen, the asymmetric active phase is much narrower in the simulations than in the theory.


Figure~\ref{fig:heatmapkave} presents analogous heatmaps for $q=0.95$ that illustrate the average node degrees calculated among nodes with the majority and the minority opinions.
In the symmetric active phase (smaller values of $p$), these average node degrees are equal to the average node degree of the simulated network $\langle k\rangle=6$, which is in accordance with the theoretical predictions.
The properties of the asymmetric active phase are less well captured.
The reason may  lie in the fluctuations that push the system into the absorbing phase more easily when $\rho$ is close to zero.

\section{Conclusions}
We analyzed the coevolving nonlinear voter model with the rewire-to-random mechanism with use of the pair approximation in which the distinction between the average degrees of nodes in different states is made.
This approach allowed us to identify two dynamically active phases---the well-known symmetric phase and the asymmetric one, which can arise from spontaneously broken symmetry.
The symmetric active phase is characterized by the same numbers of nodes in the opposite states, so none of the states is preferred in the network.
In the asymmetric active phase, on the other hand, there is a predominance of nodes in one state, so the majority opinion can be distinguished.
Only in the symmetric active phase are the average degrees of nodes in different states equal to the average node degree of the network.

In the pair approximation, for $0<q<\bar{q}$, where $\bar{q}$ depends on the average node degree of the network,
the  coevolving nonlinear voter model with the rewire-to-random mechanism exhibits only continuous phase transitions between the symmetric active and the absorbing phases.
Similar behavior is shared by the coevolving nonlinear voter model with the rewire-to-same mechanism for $0<q\leq 1$ \cite{Min:San:17}.
However, for $\bar{q}<q<1$, the pair approximation predicts much richer phase diagram for the model with the rewire-to-random mechanism than for its rewire-to-same counterpart.
In this range of the parameter, the asymmetric active phase emerges.
For $\bar{q}<q<q^*$, where $q^*=\frac{1}{6} (\sqrt{13}+1)\approx0.7676$, discontinuous phase transitions are possible, and a hysteresis loop may be observed as a  result of system bistablity.
The discontinuous phase transitions may occur between the symmetric active and the absorbing phase or directly between both active phases.
On the other hand, for $q^*\leq q<1$, two continuous phase transitions are predicted.
The first transition occurs between the symmetric and the asymmetric active phase.
At the transition point to the asymmetric active phase, the symmetry is spontaneously broken, and the majority opinion arises in the network.
Interestingly, there are different critical exponents on both sides of this transition for $q=q^*$.
As $p$ increases further, a continuous phase transition to the absorbing phase takes place.
Although the quantitative results of our approximate calculations derive from the results of Monte Carlo simulations, the appearance of the asymmetric active phase in the model was correctly predicted by the pair approximation.

In our analysis, we focused on single-layer networks.
However, since considering multi-layer networks in the coevolving nonlinear voter model with the rewire-to-same mechanism leads to the emergence of new phases \cite{Min:San:19}, the analysis of its rewire-to-random counterpart on such structures seems to be an interesting research direction.
Another interesting idea is to include links that can be in different states and consider the coevolution of node and link states \cite{Sae:San:Tor:19}.

\section*{Acknowledgments}
This work was created as a result of the research projects financed from the funds of the National Science Center (NCN, Poland) Nos. 2016/23/N/ST2/00729, 2018/28/T/ST2/00223, and 2015/19/B/ST6/02612.
The work was partially supported as RENOIR Project by the European Union Horizon 2020 Research and Innovation Program under the Marie Skłodowska-Curie Grant No. 691152, by Ministry of Science and Higher Education (Poland), Grants Nos. 34/H2020/2016 and 329025/PnH/2016, \textcolor{mycolor}{ and  by POB Research Centre Cybersecurity and Data Science of Warsaw University of Technology within the Excellence Initiative Program -- Research University (IDUB).} J.A.H. has been partially supported by the Russian Scientific Foundation, Agreement No. 17-71-30029 with co-financing of Bank Saint Petersburg.

\appendix*
\textcolor{mycolor}{
\section{Details of pair approximation}
\label{app:pairapp}
Although our network consists of undirected links, we use directed links to describe the system. 
Conceptually, this means that we replace each undirected link with two oppositely  directed links \cite{Tor:etal:17}.
Thus, the state variables are defined by the numbers of directed links connecting nodes in different states: $E_{\uparrow\uparrow}$, $E_{\uparrow\downarrow}$, $E_{\downarrow\uparrow}$, $E_{\downarrow\downarrow}$, where the first subscript corresponds to the state of a node at the origin of the link (since our network is undirected, $E_{\uparrow\downarrow}=E_{\downarrow\uparrow}$).
Now, the state variables can be expressed in the following way:
\begin{equation}
c=\frac{1}{\langle k_\uparrow\rangle N}\left(E_{\uparrow\uparrow}+E_{\uparrow\downarrow}\right),
\label{eq:cup}
\end{equation}
\begin{equation}
\rho=\frac{2}{\langle k\rangle N}E_{\uparrow\downarrow},
\label{eq:rho}
\end{equation}
and
\begin{equation}
m=\frac{1}{\langle k\rangle N}\left(E_{\uparrow\uparrow}-E_{\downarrow\downarrow}\right).
\label{eq:m}
\end{equation}
Since $\langle k\rangle N$ is the total number of directed links in the network, we have an additional constraint in the form
\begin{equation}
	E_{\uparrow\uparrow}+E_{\downarrow\downarrow}+2E_{\uparrow\downarrow}=\langle k\rangle N
	\label{eq:totalE}
\end{equation}
for our system.
Having combined Eqs.~(\ref{eq:cup}), (\ref{eq:m}), and (\ref{eq:totalE}), we get a formula for the average degree of nodes with $j=1$:
\begin{equation}
\langle k_\uparrow\rangle=\langle k\rangle\frac{1+m}{2c}.
\label{eq:kup}
\end{equation}
Similarly, we can obtain an equation for the average degree of nodes with $j=-1$:
\begin{equation}
\langle k_\downarrow\rangle=\langle k\rangle\frac{1-m}{2(1-c)}.
\label{eq:kdown}
\end{equation}
Let us denote by $\theta_j$ the conditional probability of choosing an active out-link from the out-links of a node with the opinion $j$.
In this kind of pair approximation, these probabilities are approximated by the following formulas \cite{Jed:Szn:19,Tor:etal:17,Kul:etal:18,Nar:Koz:Bar:08}:
\begin{equation}
\theta_\uparrow=\frac{E_{\uparrow\downarrow}}{E_{\uparrow\uparrow}+E_{\uparrow\downarrow}}=\frac{\rho}{1+m} \label{eq:thetaup}
\end{equation}
and
\begin{equation}
\theta_\downarrow=\frac{E_{\downarrow\uparrow}}{E_{\downarrow\downarrow}+E_{\downarrow\uparrow}}=\frac{\rho}{1-m}.\label{eq:thetadown}
\end{equation}
Having defined all the necessary quantities, we can write down three rate equations for the time evolution of our state variables. 
In the thermodynamic limit, i.e., in the limit of an infinite system size, we have
\begin{align}
\frac{d c}{d t}=&(1-p)\left[(1-c)\theta_\downarrow^q-c\theta_\uparrow^q\right],\label{eq:dc}\\
\frac{d\rho}{dt}=&\frac{2}{\langle k\rangle }\sum_{j\in\{\uparrow,\downarrow\}}c_j\theta_j^q\nonumber\\
&\times\left\lbrace(1-p)\left[\langle k_j\rangle-2q-2\left(\langle k_j\rangle-q\right)\theta_j\right]-pc_j\right\rbrace ,\label{eq:drho}
\end{align}
and
\begin{align}
\frac{dm}{dt}=&\frac{2}{\langle k\rangle}p\left[c^2\theta_\uparrow^q-(1-c)^2\theta_\downarrow^q\right]\nonumber\\
&-\frac{2}{\langle k\rangle}(1-p)\left[c\theta_\uparrow^q\langle k_\uparrow\rangle-(1-c)\theta_\downarrow^q\langle k_\downarrow\rangle\right],\label{eq:dm}
\end{align}
where we set $c_\uparrow\equiv c$ and $c_\downarrow\equiv 1-c$ to simplify notation.
}

In order to derive Eqs.~(\ref{eq:dc})-(\ref{eq:dm}), first, we find changes in $c$, $\rho$, and $m$ in one update of the model described in Sec~\ref{sec:model}.
Let us start with the changes in  $c$.
In every update, when a voter changes its opinion, this concentration increases or decreases by $1/N$. 
The opinion change is possible with probability $1-p$, and then it occurs with probability $\theta_\uparrow^q$ for nodes with $j=1$ and with $\theta_\downarrow^q$ for nodes with $j=-1$, according to the model definition.
This results in the following formula:
\begin{equation}
\Delta c=(1-p)\frac{1}{N}\left[(1-c)\theta_\downarrow^q-c\theta_\uparrow^q\right].
\end{equation}
Since $\Delta t=1/N$, taking the limit $N\to\infty$ in the above equation gives Eq.~(\ref{eq:dc}).

The changes in $\rho$ and $m$ are calculated directly from the changes in the numbers of directed links connecting nodes in different states with the use of Eqs~(\ref{eq:rho}) and (\ref{eq:m}).
However, in order to ease calculations and obtain analytical formulas, we make some approximations.
First, let us notice that when $q$ is an integer, the interaction probability, $\rho_i^q$, corresponds to the probability of choosing  with repetition $q$ disagreeing neighbors of the node $i$.
In the model without repetition, on the other hand, this probability would have the following form:
\begin{equation}
f(a_i,k_i)=\frac{\prod_{j=1}^{q}(a_i-j+1)}{\prod_{j=1}^{q}(k_i-j+1)}=\frac{a_i!(k_i-q)!}{k_i!(a_i-q)!},
\label{eq:appf}
\end{equation}
where $a_i$ and $k_i$ are the number of active links and the degree of the node $i$, respectively.
If we use in our calculations Eq.~(\ref{eq:appf}) instead of $\rho_i^q$ with the assumption that $q$ is an integer, we are able to get analytical results in a similar way to Ref.~\cite{Jed:17}, where the $q$-voter model is analyzed on static complex networks with the pair approximation.
Next, the applicability of the obtained formulas can be extended to the initial variability range of $q$.
The same procedure applied to the coevolving nonlinear voter model with the rewire-to-same mechanism leads to the formulas obtained in Ref.~\cite{Min:San:17}, where this model is analyzed.
In a similar way, one can obtain equations presented in Ref.~\cite{Rad:San:20} for the dynamics of the coevolving nonlinear voter model with the rewire-to-same mechanism and noise.
Despite such a simplification, the pair approximation that does not account for repetition captures correctly some qualitative properties of the model with repetition, such as the appearance of the asymmetric active phase in our case. 

Based on this approximate method and the model definition, we obtain the following formulas for the changes in the numbers of directed links connecting nodes in different states:
\begin{widetext}
\begin{align}
	\Delta E_{\uparrow\downarrow}=&\sum_{j\in\{\uparrow,\downarrow\}}c_j\sum_{k} P_j(k)\sum_{a=q}^{k}{{k}\choose{a}}\theta_j^{a}(1-\theta_j)^{k-a}f(a,k)\left[(1-p)(k-2a)-pc_j\right],\\
	\Delta E_{\uparrow\uparrow}=&2c\sum_{k} P_\uparrow(k)\sum_{a=q}^{k}{{k}\choose{a}}\theta_\uparrow^{a}(1-\theta_\uparrow)^{k-a}f(a,k)\left[pc-(1-p)(k-a)\right]\nonumber\\
	&+2(1-c)\sum_{k} P_\downarrow(k)\sum_{a=q}^{k}{{k}\choose{a}}\theta_\downarrow^{a}(1-\theta_\downarrow)^{k-a}f(a,k)(1-p)a,\\
	\Delta E_{\downarrow\downarrow}=&2(1-c)\sum_k P_\downarrow(k)\sum_{a=q}^{k}{{k}\choose{a}}\theta_\downarrow^a(1-\theta_\downarrow)^{k-a}f(a,k)\left[p(1-c)-(1-p)(k-a)\right]\nonumber\\
	&+2c\sum_k P_\uparrow(k)\sum_{a=q}^{k}{{k}\choose{a}}\theta_\uparrow^a(1-\theta_\uparrow)^{k-a}f(a,k)(1-p)a,
\end{align}
\end{widetext}
where $P_j(k)$ is the degree distribution associated only with nodes in the corresponding state $j\in\{\uparrow,\downarrow\}$
(since we consider an undirected network, $\Delta E_{\downarrow\uparrow}=\Delta E_{\uparrow\downarrow}$).
In the above equations, we assume that the number of active out-links, $a$, connected to the node with the degree $k$ and in the state $j$ is binomially distributed with probability $\theta_j$.
After summing over $k$ and $a$ indexes, we get the following expression for the changes in the number of active links:
\begin{align}
	\Delta E_{\uparrow\downarrow}=&\sum_{j\in\{\uparrow,\downarrow\}}c_j\theta_j^q\nonumber\\
	&\times\left\lbrace(1-p)\left[\langle k_j\rangle-2q-2\left(\langle k_j\rangle-q\right)\theta_j\right]-pc_j\right\rbrace,
\end{align}
On the other hand, the numbers of inactive links change in the following way:
\begin{align}
	\Delta E_{\uparrow\uparrow}=&2c\theta_\uparrow^q\left[pc-(1-p)\left(\langle k_\uparrow\rangle-q\right)\left(1-\theta_\uparrow\right)\right]\nonumber\\
	&+2(1-c)(1-p)\theta_\downarrow^q\left[q+\left(\langle k_\downarrow\rangle-q\right)\theta_\downarrow\right],\\
	\Delta E_{\downarrow\downarrow}=&2(1-c)\theta_\downarrow^q\left[p(1-c)-(1-p)\left(\langle k_\downarrow\rangle-q\right)\left(1-\theta_\downarrow\right)\right]\nonumber\\
	&+2c(1-p)\theta_\uparrow^q\left[q+\left(\langle k_\uparrow\rangle-q\right)\theta_\uparrow\right].
\end{align}
Equations~(\ref{eq:drho}) and (\ref{eq:dm}) result directly from the above equations and the definitions of $\rho$ and $m$, i.e., Eqs.~(\ref{eq:rho}) and (\ref{eq:m}).
The obtain results depend only on the average node degree of the network.
A similar situation arises in the case of the pair approximation applied to the $q$-voter model \cite{Jed:17} and the noisy threshold $q$-voter model \cite{Vie:etal:20} considered without repetition on static networks.
However, taking into account repetition in these models leads to the appearance of other moments in the solutions \cite{Per:etal:18,Vie:etal:20}.

\bibliography{12112019_literature}

\end{document}